\documentclass[letter]{aa}  

\usepackage{graphicx}
\usepackage{txfonts}
\usepackage[hidelinks,colorlinks=true,citecolor=cyan]{hyperref}
\usepackage{isotope}

\newcommand{\msun}{$\rm M_{\odot}$}
\newcommand{\p}{$p$}

\newcommand{\s}{$s$}
\newcommand{\g}{$\gamma$}
\newcommand{\A}{$\alpha$}
\newcommand{\fz}{$\rm F_0$}
\newcommand{\fg}{$\rm F_{\gamma}$}

\newcommand{\fx[1]}{$\rm F_{#1}$}

\graphicspath{{./}{figures/}}

\begin{document}

   \title{The \g-process nucleosynthesis in core-collapse supernovae}
   \subtitle{II. Effect of the explosive recipe}

    \author{
        L. Roberti \inst{1,2,3,4}
        \and
        M. Pignatari \inst{1,2,5,4}
        \and 
        C. Fryer \inst{6}
        \and 
        M. Lugaro \inst{1,2,7,8}
    }
    
    \institute{Konkoly Observatory, Research Centre for Astronomy and Earth Sciences, HUN-REN, Konkoly Thege Mikl\'{o}s \'{u}t 15-17, H-1121 Budapest, Hungary; 
    \and
                CSFK, MTA Centre of Excellence, Budapest, Konkoly Thege Miklós út 15-17, H-1121, Hungary
    \and
                INAF -- Osservatorio Astronomico di Roma Via Frascati     33, I-00040, Monteporzio Catone, Italy
    \and            
                NuGrid Collaboration, \url{http://nugridstars.org}
    \and
               E. A. Milne Centre for Astrophysics, University of Hull, Hull HU6 7RX, UK
    \and
                Center for Theoretical Astrophysics, Los Alamos National Laboratory, Los Alamos, NM, 87545 USA
    \and
               E\"otv\"os Lor\'and University, Institute of Physics and Astronomy, Budapest 1117, P\'azm\'any P\'eter s\'er\'any 1/A, Hungary
   \and
            School of Physics and Astronomy, Monash University, VIC 3800, Australia}

   \date{Received March 15, 2024; accepted May 7, 2024}

 
  \abstract
   {The \g\  process in core-collapse supernovae (CCSNe) can produce a number of neutron-deficient stable isotopes heavier than iron (\p\ nuclei). 
   However, current model predictions do not fully reproduce solar abundances, especially for \isotope[92,94]{Mo} and \isotope[96,98]{Ru}.
   }
   {We investigate the impact of different explosion energies and parametrizations on the nucleosynthesis of \p\  nuclei, by studying stellar models with different initial masses and different CCSN explosions.
   }
   {We compared the \p-nucleus yields obtained using a semi-analytical method to simulate the supernova to those obtained using hydrodynamic models. We explored the effect of varying the explosion parameters on the \p--nucleus production
   in two sets of CCSN models with initial masses of 15, 20, and 25 \msun\ at solar metallicity. We calculated a new set of 24 CCSN models (eight for each stellar progenitor mass) and compared our results with another recently published set of 80 CCSN models that includes a wide range of explosion parameters: explosion energy or initial shock velocity, energy injection time, and mass location of the injection.  
   }
   {We find that the total \p-nucleus yields are only marginally affected by the CCSN explosion prescriptions if the \g-process production is already efficient in the stellar progenitors due to a C–O shell merger.
   In most CCSN explosions from progenitors without a C-O shell merger, the \g-process yields increase with the explosion energy by up to an order of magnitude, depending on the progenitor structure and the CCSN prescriptions. The general trend of the \p-nucleus production with the explosion energy is more complicated if we look at the production of single \p\   nuclei. The light \p-nuclei tend to be the most enhanced with increasing explosion energy. In particular, for the CCSN models where the \A-rich freeze-out component is ejected, the yields of the lightest \p\   nuclei (including \isotope[92,94]{Mo} and \isotope[96]{Ru}) increase by up to three orders of magnitude.  
   }
   {We provide the first extensive study using different sets of massive stars of the impact of varying CCSN explosion prescriptions on the production of \p\   nuclei. Unlike previous expectations and recent results in the literature, we find that the average production of \p\  nuclei tends to increase with the explosion energy. We also confirm that the pre-explosion production of \p\ nuclei in C-O shell mergers is a robust result, independent of the subsequent explosive nucleosynthesis. More generally, a realistic range of variations in the evolution of stellar progenitors and in the CCSN explosions might boost the CCSN contribution to the galactic chemical evolution of \p\ nuclei.
   }


   \maketitle
%

\section{Introduction} \label{sec:intro}

    Massive stars (with initial masses > 8--10 \msun) end their lives with the formation and the collapse of a Fe core. The collapse halts and bounces back as soon as the central infalling material reaches the typical nuclear density. The bounce of the innermost regions of the star generates a shock wave that, in some cases, is able to emerge from the Fe core and propagate into the envelope of the star \citep{burrows:21}. The shock wave induces local compression and heating, and this causes nucleosynthesis to occur within a few seconds rather than the days or even years it takes during the pre-supernova evolution. This explosive nucleosynthesis alters the chemical composition produced during the hydrostatic evolution \citep[see, e.g.,][]{rauscher:02,pignatari:16,LC18}. The most abundant nuclear species in compositions of this so-called core-collapse supernova (CCSN) ejecta are the Fe-peak and \A\ elements, mostly synthesized through (quasi)statistical equilibrium processes in the explosion, together with the stellar burning products from the hydrostatic evolution of the star \citep{woosley:95}.

    Core-collapse supernovae are also one of the production sites of the so-called \p\  nuclei via the \g\  process \citep[][]{rayet:95,arnould:03,pignatari:16a,roberti:23a}. The ONe-rich layers in massive stars during a CCSN provide the necessary conditions (i.e., high temperatures and abundances of trans-iron seeds) to synthesize the \p\  nuclei via a chain of photo-disintegrations  \citep{woosley:78}. Several discrepancies still arise when comparing theoretical model predictions to each other and with solar abundances \citep{roberti:23a}. In particular, (i) the production is too low to match the solar abundances; (ii) the isotopic ratios of theoretical yields produced in the same temperature range cannot reproduce the solar ratios; and (iii) \isotope[92,94]{Mo} and \isotope[96,98]{Ru} show an even more severe underproduction compared to the other \p\  nuclei \citep{rauscher:13,pignatari:16a}. While additional nucleosynthetic processes and astrophysical phenomena have been invoked to explain the solar abundances of \p\ nuclei (see \citealt{roberti:23a} for a recent overview of the production sites of \p\ nuclei), this remains an open problem. For example, C--O shell mergers in the late stage of massive star evolution can play a crucial role in boosting the final yields of \p\ nuclei, but this does not explain the underproduction of the Mo and Ru isotopes \citep{rauscher:02,ritter:18a}.

    \cite{choplin:22} show that fast rotation in a 25 \msun\ star at subsolar metallicity can enhance the efficiency of the \g-process nucleosynthesis, due to a higher production of trans-iron seeds via the slow neutron capture nucleosynthesis (the \s\  process).
    The possibility of including such results in models of galactic chemical evolution needs to be explored as it may modify our current understanding, that the contribution from previous generations of massive stars to the Solar System abundances of \p\ nuclei is negligible \citep{travaglio:18}.
    \cite{choplin:22} also briefly discuss how the different explosion energies affect the production of \p\ nuclei for a wide range of energies (0.3 -- 100 foe). They conclude that, at least in the specific case of their 25 \msun\ progenitor, the explosion energy has only a marginal impact on the total yields of \p\ nuclei because the only effect of increasing the energy is the shift of the \g-process zone outward in mass. Nevertheless, a detailed study of the dependence of \g-process nucleosynthesis, and of the production of \p\ nuclei in general, on different explosion energies and explosive prescriptions is still missing. With this work we aim to start an exploration in this direction.

    In the first paper of this series \citep[][hereafter Paper I]{roberti:23a}, we presented the first step of our study of \g-process nucleosynthesis in CCSNe. We analyzed and compared the yields from five different sets of massive star models from the literature. In this second work, we discuss the effect of different explosive prescriptions on 15, 20, and 25 \msun\ star progenitors. In Sect. \ref{sec:datasets} we describe the different explosive prescriptions; Sect. \ref{sec:gamma} summarizes the conditions that affect the \g-process nucleosynthesis. In Sect. \ref{sec:f0} we discuss how the different explosion prescriptions influence the average overproduction factors for \p\ nuclei, and in Sect. \ref{sec:conclusions} we discuss and summarize our results.


\section{Datasets} \label{sec:datasets}

    We explored the effect of different supernova prescriptions on the \g\  process and on the \p-nucleus yields in the two set of models described below, with initial masses of 15, 20, and 25 \msun\ and solar metallicity. We considered a broad range of explosion possibilities using different parameters that describe both the explosion energy and the amount of material that is ejected (versus the amount that is incorporated onto the compact object). Multiple parameters are currently required to capture the uncertainties in the explosive engine.  The explosion energy depends on the growth time of the convection \citep{2021ARep...65..937F,2022ApJ...931...94F,boccioli:23} as well as on the existence of energy sources, such as continued accretion after the shock is launched  (especially if there is sufficient angular momentum to form an accretion disk) or the development of a magnetar~\citep{2021MNRAS.508.5390S,2022ApJ...935..108S}.  The mass cut also depends on the convective growth time~\citep{2022ApJ...931...94F}.  Fallback (which also depends on the explosion energy) can also impact the final remnant mass and can be particularly important in asymmetric explosions~\citep{2009ApJ...699..409F,fryer:12}.  Our two supernova prescriptions help cover the broad uncertainties in the ejecta from supernovae.

    \subsection{Sedov blast-wave (SBW) explosion } \label{subsec:rit}

        \begin{table}[!t]
           \caption{Selected features of the CCSN models calculated using the SBW approach.
           }
           \label{tab:sbw}      
           \centering                          
           \begin{tabular}{crcrrrrrrc}        
           \hline\hline                 
           M$_{\rm ini}$            &  M$_{\rm rem}$ & v$_{\rm shock}$      & Si(c) & Si(i) & O     & Ne    &  C    \\ 
           \hline
                           &       & 0.2              & -     & -     & -     & -     & -     \\
                           &       & 0.4              & -     & -     & -     & 1.64  & 1.67  \\
                           &       & 1                & -     & -     & 1.64  & 1.78  & 1.82  \\
                           &       & 1.33             & -     & 1.63  & 1.68  & 1.85  & 1.90  \\
           15              & 1.61  & 2                & 1.63  & 1.68  & 1.74  & 1.96  & 2.03  \\
                           &       & 3                & 1.68  & 1.75  & 1.82  & 2.11  & 2.20  \\
                           &       & 4                & 1.72  & 1.80  & 1.90  & 2.24  & 2.34  \\
                           &       & 5                & 1.73  & 1.82  & 1.91  & 2.23  & 2.33  \\
           \hline     
                           &       & 0.2              & -     & -     & -     & -     & -     \\   
                           &       & 0.4              & -     & -     & -     & -     & -     \\
                           &       & 1                & -     & -     & -     & 3.01  & 3.23  \\
                           &       & 1.33             & -     & -     & -     & 3.32  & 4.80  \\
           20              & 2.73  & 2                & -     & -     & 2.79  & 4.75  & 4.80  \\
                           &       & 3                & -     & 2.81  & 3.22  & 4.75  & 4.80  \\
                           &       & 4                & -     & 3.11  & 3.57  & 4.75  & 4.80  \\
                           &       & 5                & 2.85  & 3.28  & 3.76  & 4.75  & 4.80  \\
           \hline    
                           &       & 0.2              & -     & -     & -     & -     & -     \\  
                           &       & 0.4              & -     & -     & -     & -     & -     \\
                           &       & 1                & -     & -     & -     & -     & -     \\
                           &       & 1.33             & -     & -     & -     & -     & -     \\
           25              & 5.71  & 2                & -     & -     & -     & -     & 6.79  \\
                           &       & 3                & -     & -     & -     & 6.60  & 6.79  \\
                           &       & 4                & -     & -     & -     & 6.60  & 6.79  \\
                           &       & 5                & -     & -     & -     & 6.60  & 6.79  \\
           \hline          
           \end{tabular}
           \tablefoot{M$_{\rm ini}$, M$_{\rm rem}$, and v$_{\rm shock}$ are the initial mass of the progenitor, the remnant mass, and the initial shock velocity, respectively. The last five columns indicate the outer mass coordinate corresponding to the zones exposed to the different explosive burning stages: complete Si burning ($\rm T_9>5$, "Si(c)"), incomplete Si burning ($5>\rm T_9>4$, "Si(i)"), explosive O burning ($4>\rm T_9>3.3$, "O"), explosive Ne burning ($3.3>\rm T_9>2.1$, "Ne"), and explosive C burning ($2.1>\rm T_9>1.9$, "C"); here,  $\rm T_9 = T(K)/10^9 K$. The symbol "-" means that no zone is exposed to the indicated explosive burning process. All masses are in \msun\ units, and the initial shock velocity is in units of $\rm 10^9\ cm\ s^{-1}$.}\\
        \end{table}
        
        We simulated a new set of CCSN explosions using the same approach and the same three massive star models calculated with the \verb|MESA| code \citep[][and references therein]{paxton:15} by \cite{ritter:18} at solar metallicity \citep[Z=0.02, based on][]{grevesse:93}. We varied the initial shock velocity within the range $2\times10^8 \rm cm/s$ to $5\times10^9 \rm cm/s$, which correspond to a factor of 10 below and 2.5 above the \citeauthor{ritter:18} standard supernova explosion setup. We briefly recall here the main features of the explosive prescription. The supernova explosion was modeled using a semi-analytical approach, with the prescription for the remnant mass from \cite{fryer:12} and the \cite{sedov:46} blast wave (SBW) solution for the determination of the peak velocity of the shock throughout the stellar structure. The temperature and the density in each layer of the star after the shock were then obtained by imposing a strong shock limit \citep{chevalier:89} and the radiation-dominated shock \citep[Eqs. 3--5 of][]{pignatari:16a}. After the passage of the shock, the temperature and the density decrease with time as a consequence of the cooling of the shocked material. This temporal evolution was described using a variant of the adiabatic exponential decay \citep{hoyle:64,fowler:64}. As in \cite{ritter:18}, we set an artificial cap to the maximum velocity of $5\times10^9 \rm cm/s$ to simulate the deceleration due to viscous forces. We note that in this set we kept the remnant mass fixed using the prescription from \cite{fryer:12} and decided to study the effect on the \g-process nucleosynthesis of the explosion energy only.
        
        Both the hydrostatic and explosive nucleosynthesis were calculated using the same version of the JINA Reaclib database used by \cite{ritter:18} \citep[V1.1;][]{cyburt:11} and with an updated version of the Multi-zone Post-Processing Network -- Parallel (MPPNP) code \citep[][Paper I]{Pignatari:2012dw,pignatari:16a,ritter:18}. \tablename~\ref{tab:sbw} presents a summary of the main features of these models.
        
    \subsection{Lawson set (LAW)} \label{subsec:law}

        We further used the CCSN yields from the models of \cite{jones:19}, \cite{andrews:20}, and \citet[hereafter LAW]{lawson:22} to study how different supernova parameterizations affect the production of \p\ nuclei. The progenitors were computed with the 1D {\scshape Kepler} code \citep{heger:10} using the solar metallicity from \cite{grevesse:93}, Z=0.02. The explosions were calculated using a 1D hydrodynamic code that mimics the 3D convection-enhanced supernova engine \citep{fryer:18}. \cite{fryer:18} performed a parametric study of the explosion energy, compact remnant production, and basic nucleosynthetic yields of CCSNe by varying the power ($E_{\rm inj}$), duration ($t_{\rm inj}$), and extent of the energy injection region ($M_{\rm inj}$) in the stellar progenitors. On the basis of this structure and evolution, \cite{jones:19}, \cite{andrews:20}, and \cite{lawson:22} calculated the hydrostatic nucleosynthesis using the MPPNP code and the explosive nucleosynthesis using the Tracer particle Post-Processing Network – Parallel (TPPNP) code \citep[][]{jones:19}. Both the hydrostatic and explosive nucleosynthesis were calculated using version V2.2 of the JINA Reaclib database. These calculations consist in a set of yields from 80 different explosive parametrizations, including 23 explosions of the 15 \msun\ model, 31 of the 20 \msun\ model, and 26 of the 25 \msun\ model \footnote{The full datasets of these models can be found at \url{https:// ccsweb.lanl.gov/astro/nucleosynthesis/nucleosynt hesis_astro.html}}. The $E_{\rm inj}$ ranged between 3 and 200 foe, $M_{\rm inj}$ between 0.02 and 0.2 \msun, and $t_{\rm inj}$ between 0.1 and 1 s. They produce explosion energies and remnant masses in the interval between 0.34 and 18.4 foe and 1.5 and 5.6 \msun, respectively.
        


        

\section{The \g\ process and the CCSN explosion} \label{sec:gamma}

    \begin{figure}[!t]
        \centering
        \includegraphics[width=\linewidth]{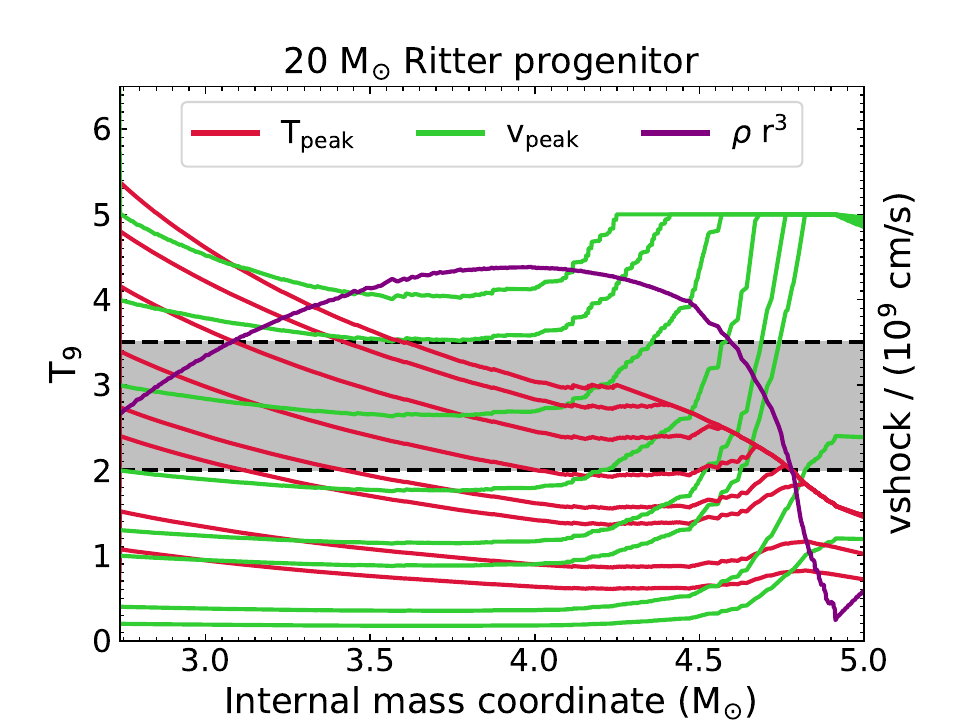}
        \caption{Peak shock temperature (solid red lines) and velocity (solid green lines) for different initial explosive conditions in the 20 \msun\ progenitor from \cite{ritter:18}. The purple line represents the logarithm of the quantity $\rho r^3$ in the pre-supernova model in arbitrary units. The gray region represents the temperature range for the activation of the \g\  process, i.e., between 2 GK and 3.5 GK. Note that the saturation of the shock velocity at $\sim$ 4.2 \msun\ at the value of $\rm 5\times 10^9\ cm\ s^{-1}$ is due to the artificial cap imposed for viscous forces (see Sect. \ref{subsec:rit}).}
        \label{fig:tpeak}
    \end{figure}
    
    The temperature required to produce \p\ nuclei via the \g\  process ranges between 2 and 3.5 GK \citep[see, e.g.,][]{pignatari:16a}, which in CCSNe roughly corresponds to the typical temperature of explosive Ne burning \citep{limongi:08}. Based on this temperature range, we defined the "hot" \g-process component as the light \p\ nuclei that are produced at a temperature near $\sim3.5$ GK (for example, \isotope[74]{Se} and \isotope[102]{Pd}) and the "cold" \g-process component as the intermediate and heavy \p\ nuclei that are produced at a temperature near $\sim2$ GK (for example, \isotope[130]{Ba} and \isotope[196]{Hg}). 

    \begin{figure*}[!t]
        \centering
        \includegraphics[scale=0.45]{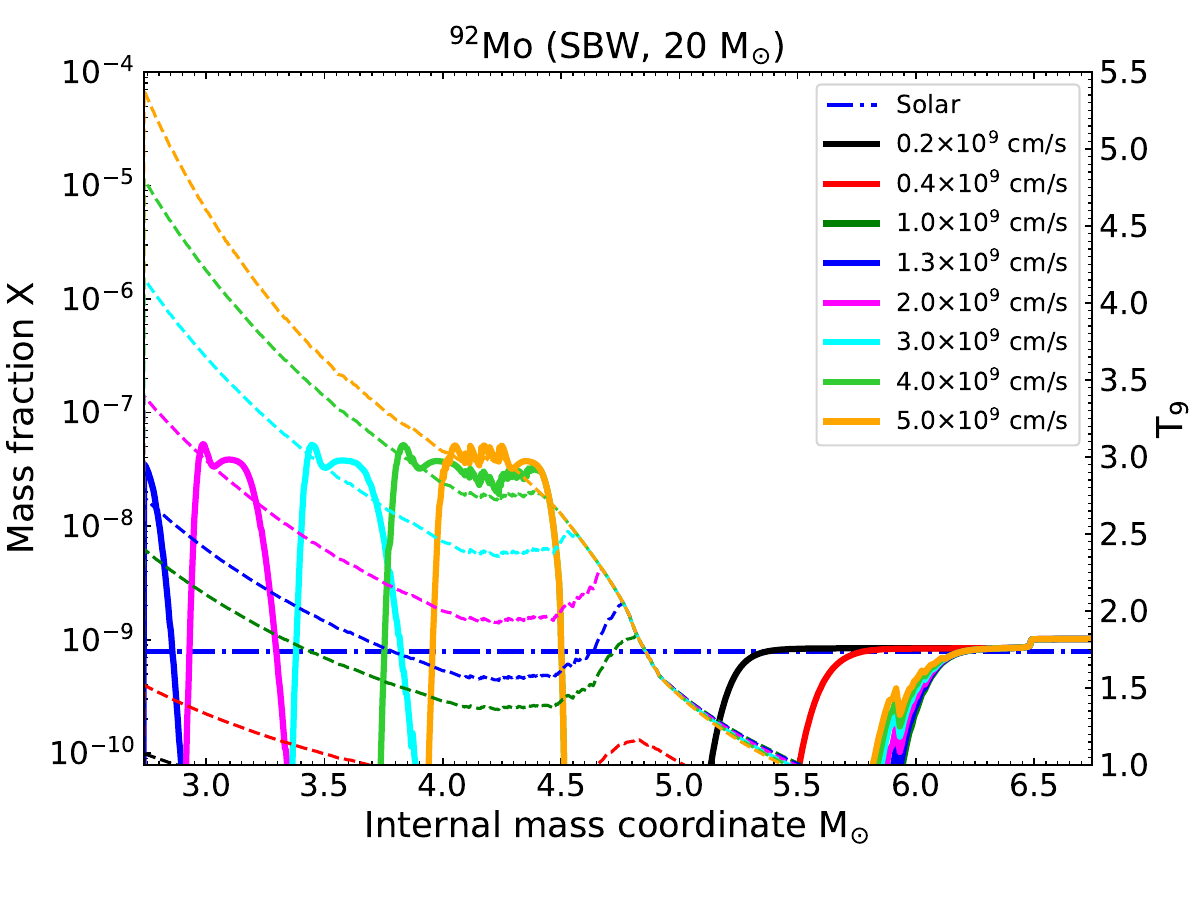}
        \includegraphics[scale=0.45]{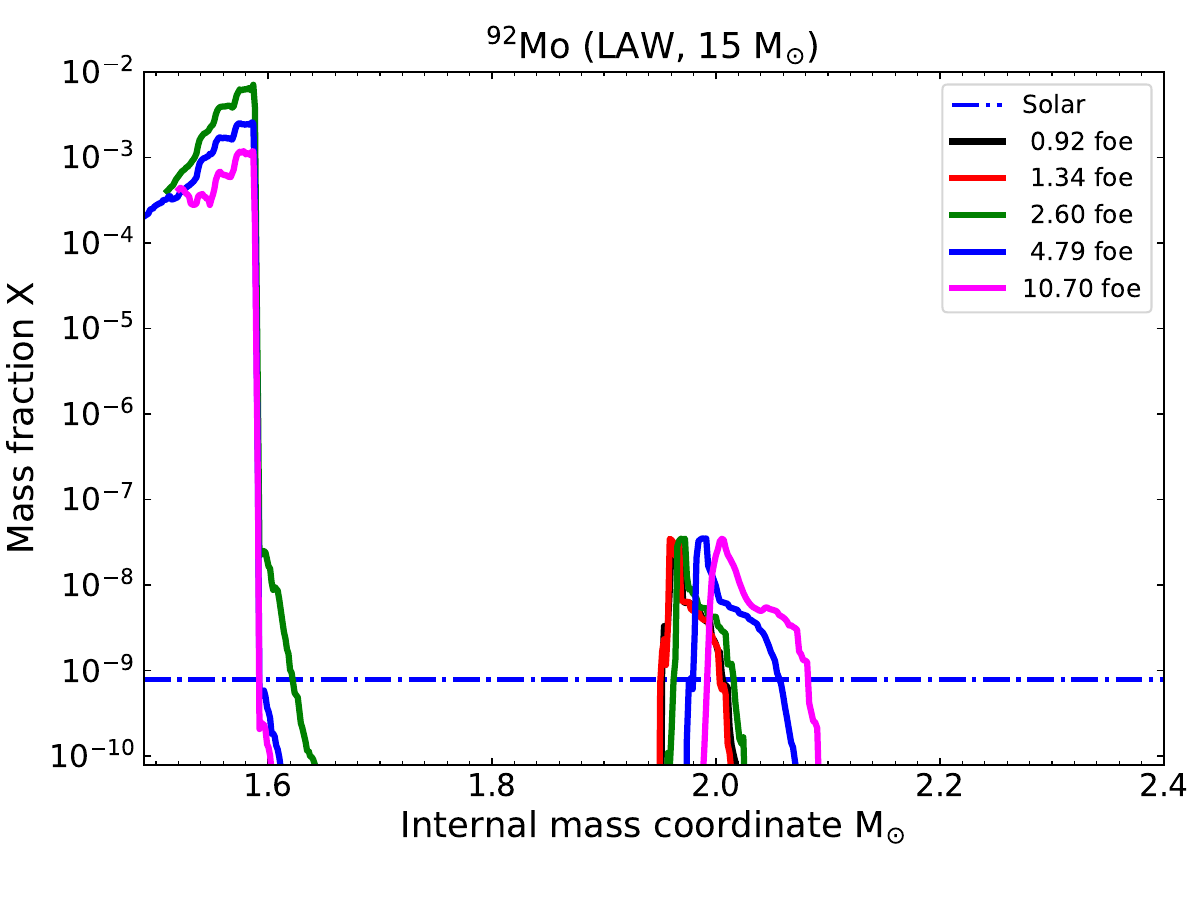}
        \includegraphics[scale=0.45]{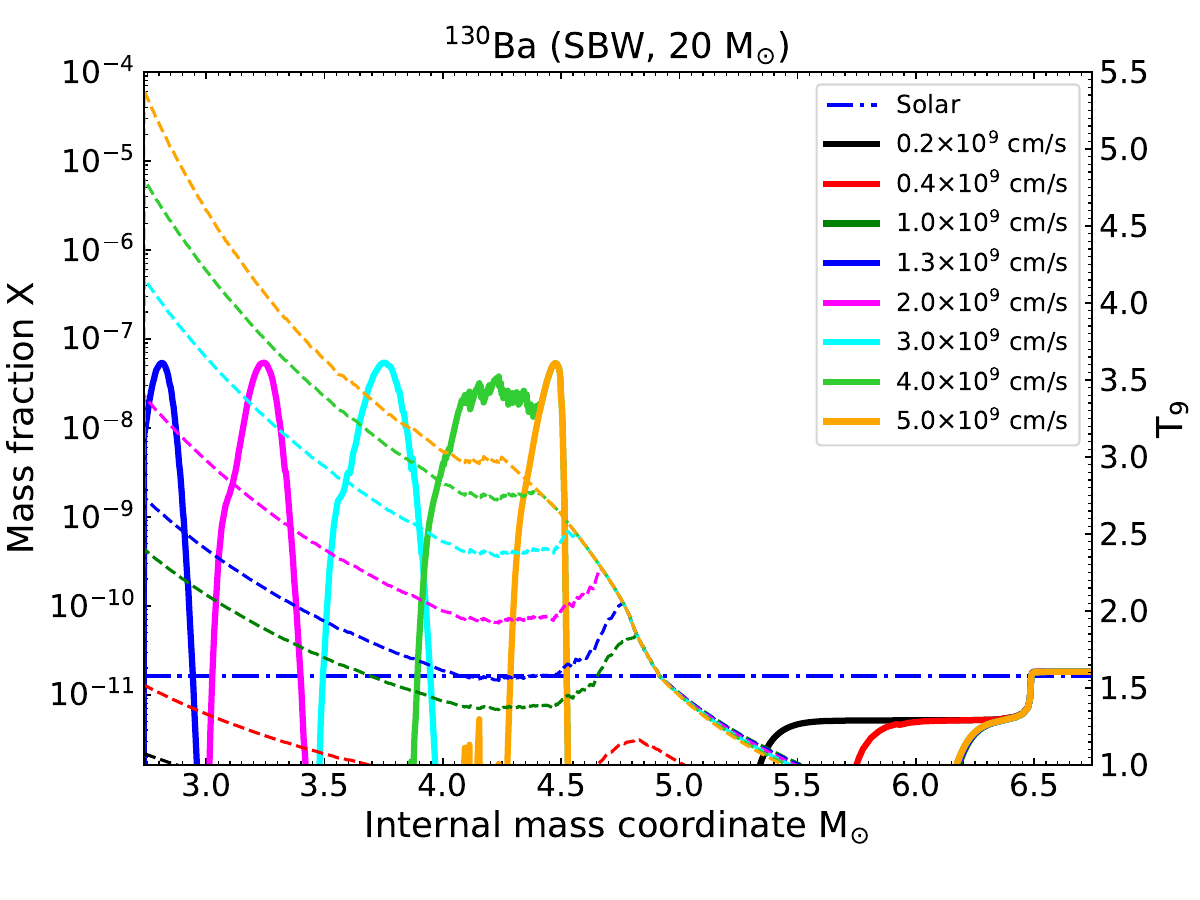}
        \includegraphics[scale=0.45]{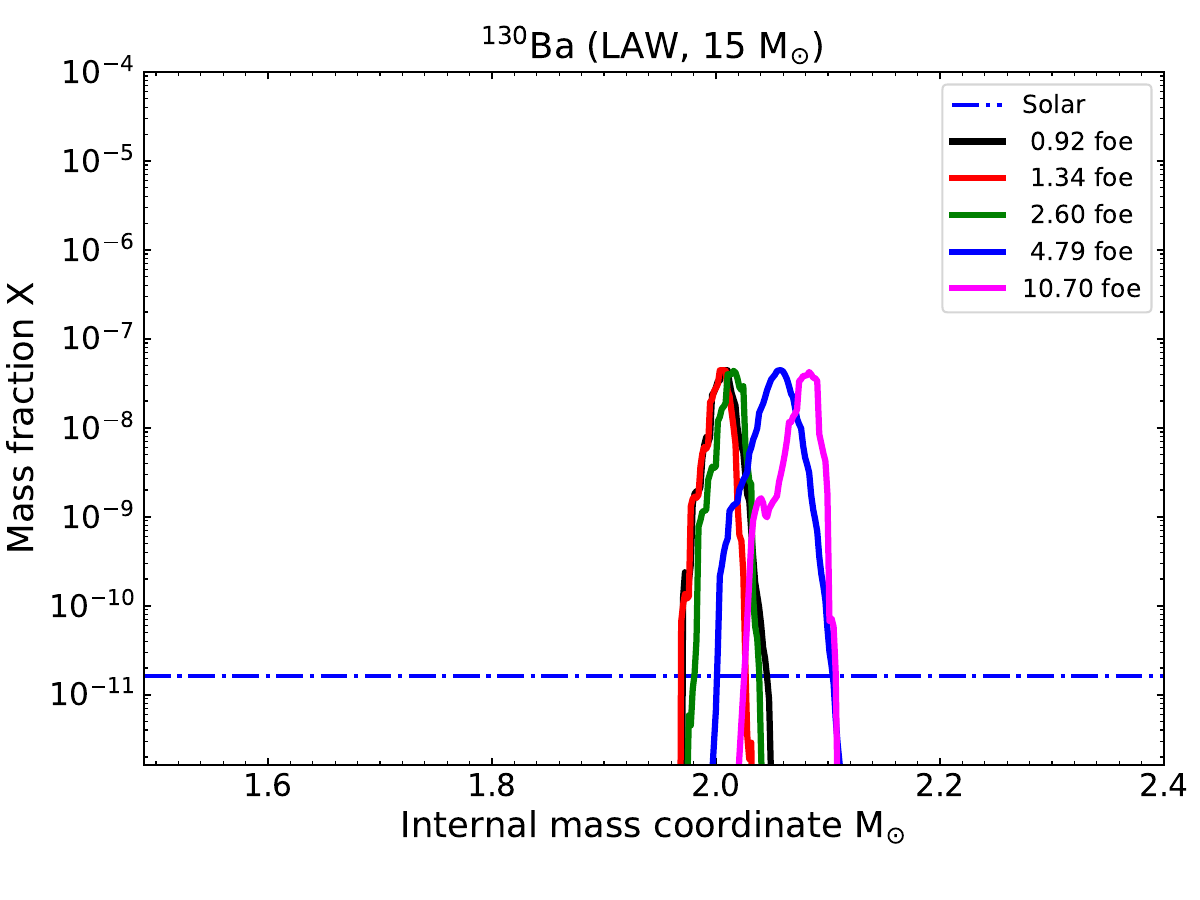}
        \includegraphics[scale=0.45]{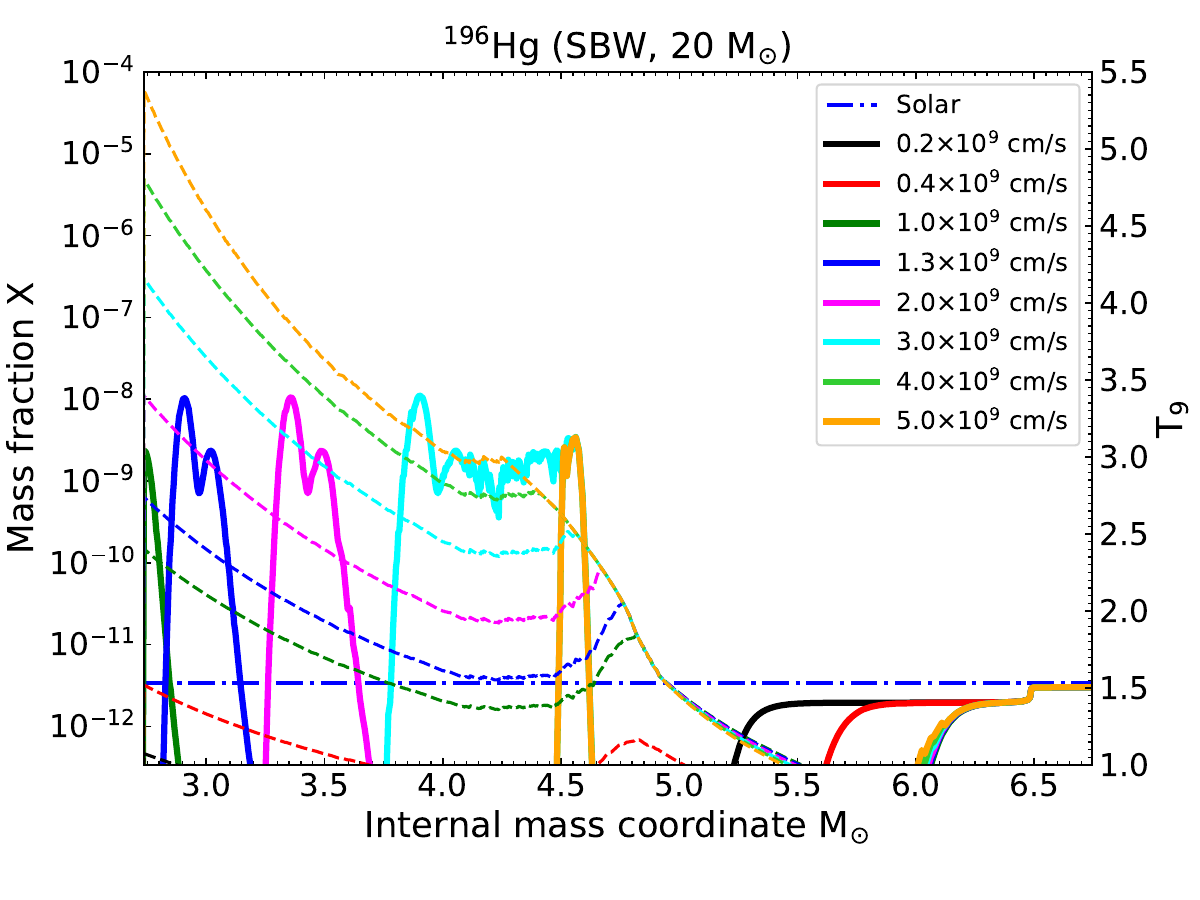}
        \includegraphics[scale=0.45]{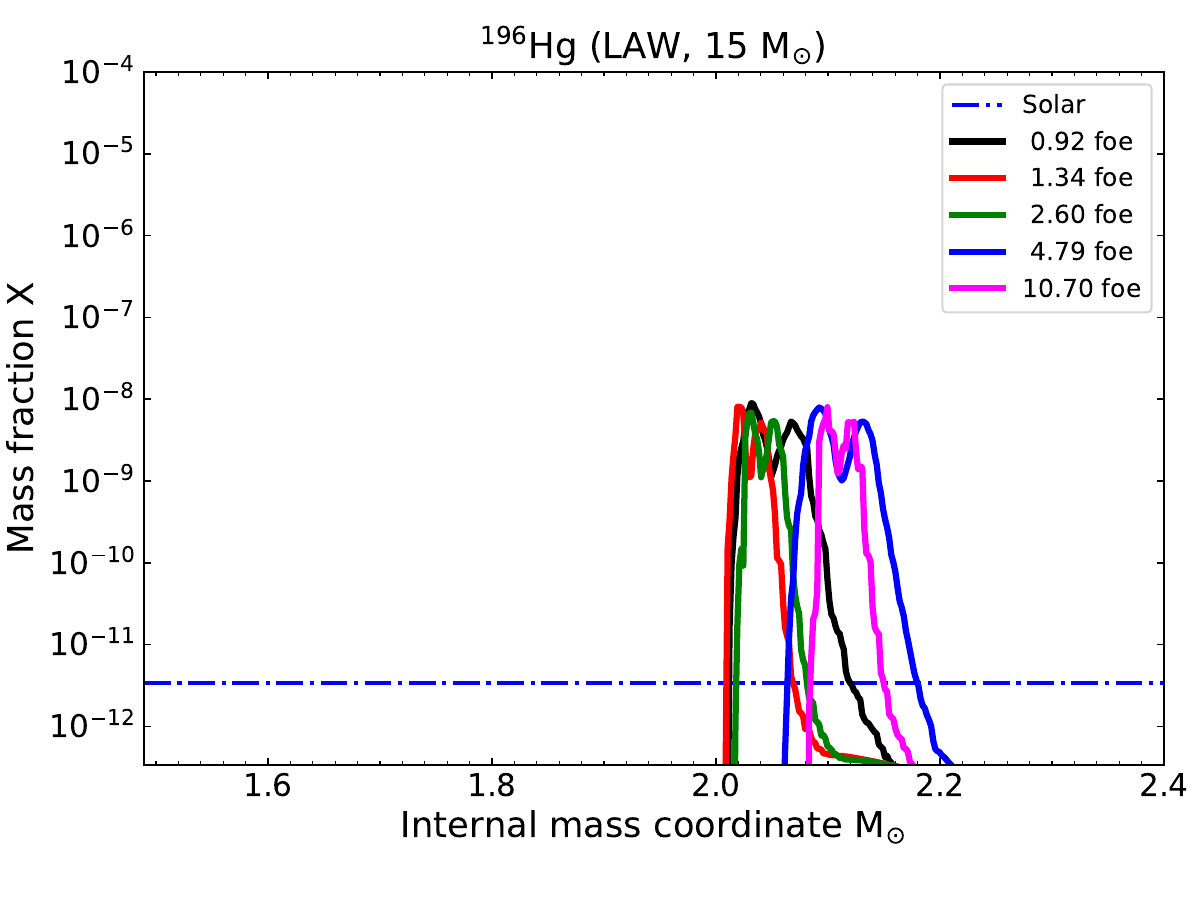}
        \caption{Effect of different explosion energies and prescriptions on three representative p–nuclei. Left panels: \isotope[92]{Mo} (upper panel), \isotope[130]{Ba} (central panel), and \isotope[196]{Hg} (lower panel) abundances in mass fractions as a function of the internal mass coordinate (solid lines, left axis) and explosion peak temperature in $\rm T_9$ (dashed lines, right y-axis) in the 20 \msun\ SBW models. Different colors denote different shock velocities. Right panels: Same as the left panels but for a sample of the 15 \msun\ LAW models, without the explosion peak temperature plotted. The horizontal dashed-dotted  blue line in each panel represents the solar abundance of the representative isotope in mass fractions taken from \cite{asplund:09}.}
        \label{fig:spag}
    \end{figure*}
    
    The main effect of adopting different explosion prescriptions is to shift the mass coordinate in the structure of the supernova progenitor where the shock wave reaches a certain velocity and, therefore, a certain temperature. Consequently, a basic assumption that has often been made is that the production of \p\ nuclei is almost independent of the properties of the explosion, since a different propagation of the shock would, in principle, only shift the location mass of the explosive Ne burning \citep[see][and our Table \ref{tab:sbw}]{choplin:22}. 
    However, according to the Sedov solution\footnote{From the Sedov solution: $E_{\rm exp} \propto \rho\ r^5\ t^{-2} \rightarrow v_{\rm shock} \propto (\rho\ r^3)^{-1}$.}, a local decrease in the quantity $\rho r^3$ would facilitate the local propagation of the shock wave and its acceleration \citep[][]{sedov:46,woosley:02}.
    The acceleration results in a temperature increase, which in some cases (especially for high-energy explosions) can lead to the temperature threshold being exceed again  for the \g\  process and can increase the extent in mass of the zone where the \p\ nuclei are produced. Figure \ref{fig:tpeak} shows this occurrence in the case of a 20 \msun\ model from the SBW set. In the case of the most energetic explosions, the peak temperature spends more time in the \g-process area (represented by the gray band in the figure) due to the acceleration of the shock wave where $\rho r^3$ decreases. The left panels of Fig. \ref{fig:spag} show the effect of the late acceleration of the shock wave (in the same model as \figurename~\ref{fig:tpeak}) on the final abundances of three representative \p\ nuclei (\isotope[92]{Mo}, \isotope[130]{Ba}, and \isotope[196]{Hg}). The abundance peaks corresponding to the \p\ nuclei are progressively more extended in mass with increasing initial shock velocity, at least in the models with $\rm v_{shock} \geq 2\times10^9 cm/s$. However, in the most energetic explosions, the abundance of all the \p\ nuclei drops abruptly at $\sim4.5$ \msun. This mass coordinate corresponds to the interface between the CO and He cores, where the temperature and the density of the shock sharply decrease. If the temperature is still high enough for an efficient \g-process nucleosynthesis when the shock wave reaches the CO--He core interface, the production of the cold \g-process components are generally suppressed because the peak temperature decrease is too sharp to extend farther into the \g-process region. We observe a qualitatively similar, albeit less intense, behavior in the LAW models (right panels of Fig. \ref{fig:spag}). In this case, the interplay among the multiple explosion parameters often leads to a non-monotonic increase in the \g-process yields with the final explosion energy.

    The choice of mass cut (i.e., the mass coordinates that separate the ejecta from the remnant mass) plays a relevant role in the ejection of \p\ nuclei. A more external mass cut will (partly) lock the ashes of the explosive Ne burning within the remnant; conversely, a deeper mass cut will allow the supernova to fully eject the \g-process products and, in some cases, to eject some material rich in the lightest \p\ nuclei (\isotope[74]{Se}, \isotope[78]{Kr}, \isotope[84]{Sr}, \isotope[92-94]{Mo}, and, possibly, \isotope[96-98]{Ru}) produced by the \A-rich freeze-out \citep[see, e.g.,][]{woosley:02,rauscher:13,pignatari:16a,lugaro:16}. The upper right panel of Fig. \ref{fig:spag} shows the large abundances of the isotope \isotope[92]{Mo} synthesized in the \A-rich freeze-out in some of the 15 \msun\ models from the LAW set. To the contrary, no significant \A-rich freeze-out component is ejected in any of the SBW models. We recall that in the SBW models the mass cut is independent of the explosion energy because it is predetermined by the adopted semi-analytical prescription, while for LAW models it is the result of the solution of the hydrodynamic equations.
    
    In conclusion, different choices regarding the explosion prescription can significantly alter the amount of \p-nucleus-rich material ejected by the CCSN. We analyze this effect in more detail in the next section. 

\section{Discussion: Effect of different explosion prescriptions on the production of \p\  nuclei} \label{sec:f0}

    We defined the overproduction factor, $\rm F_i$, of a certain isotope, $i$, as the ratio between its mass fraction, $X_{\rm i}$ (which is equal to the total integrated yield divided by the total mass ejected) and the solar mass fraction, $X_{\rm i,\odot}$, from \cite{asplund:09}. In Paper I we discussed the use of the average overproduction factor, \fz, defined as $\rm F_0 = (\sum_{i=1}^{35} F_i)/35$, as a proxy for the overall \p-nucleus production \citep[see, e.g.,][]{arnould:03,pignatari:13}. We also introduced alternative definitions of this parameter to disentangle the \g\  process from the other processes that can contribute to the production of \p\ nuclei. In particular, we defined \fg\ as the average overproduction factor of the three most produced \g-only nuclei. Since \fg\ reflects the dominant \g-process production, we can consider the isotopes contributing to \fg\ as "thermometers" for the \g\   process and use them to characterize the differences between the explosive prescriptions. To approximately estimate the contribution from CCSNe to the solar abundances of the \p\ nuclei and to discuss how the \p\ nucleus synthesis changes for different models, we compared \fz\ and \fg\ with the overproduction of \isotope[16]{O} (\fx[O16]), the most abundant metal ejected from massive stars. The \fz, \fg, and \fx[O16] of each model are given in Tables \ref{tab:15}, \ref{tab:20}, and \ref{tab:25}.
        
%

    \begin{figure*}[!t]
        \centering
        \includegraphics[width=0.48\textwidth]{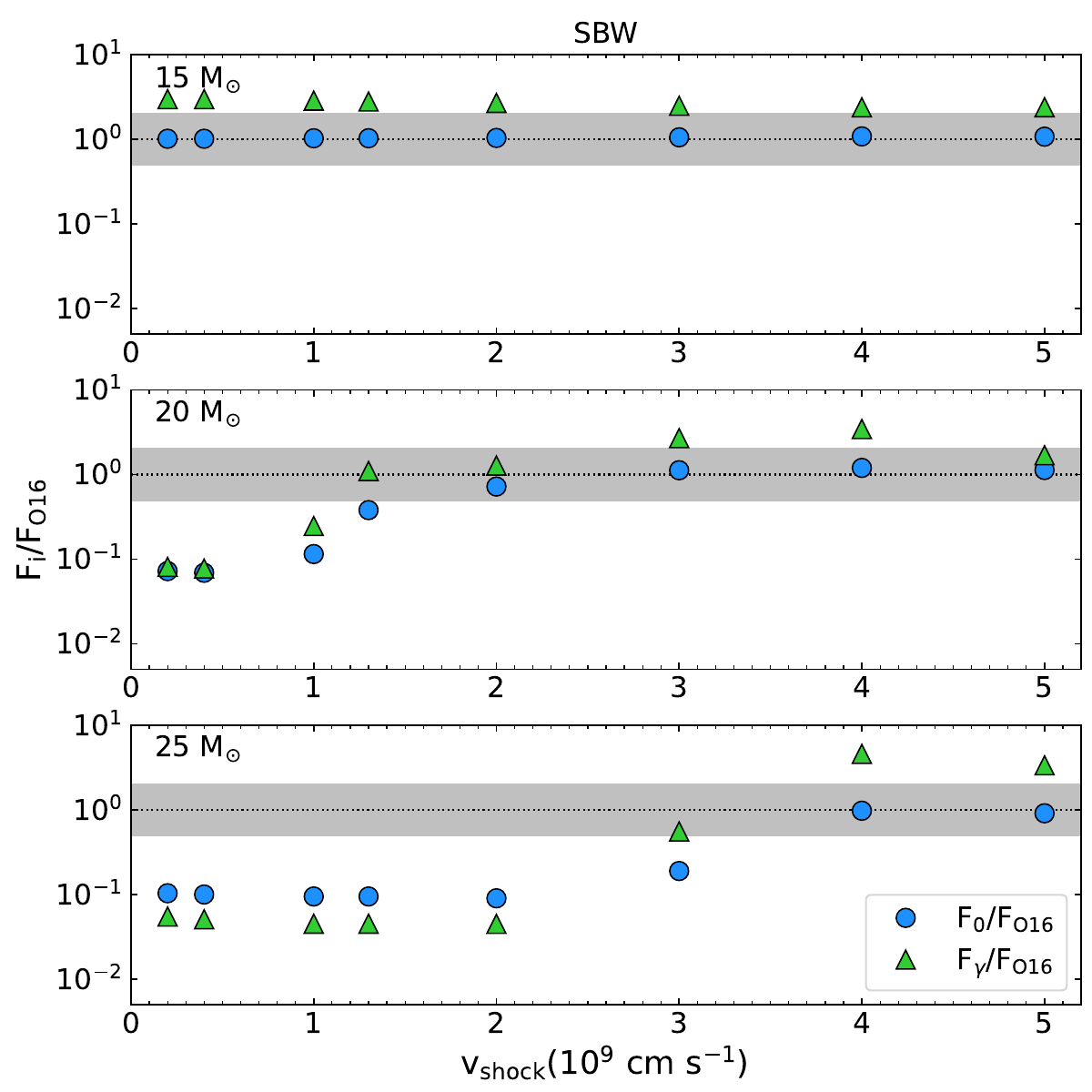}
        \includegraphics[width=0.48\textwidth]{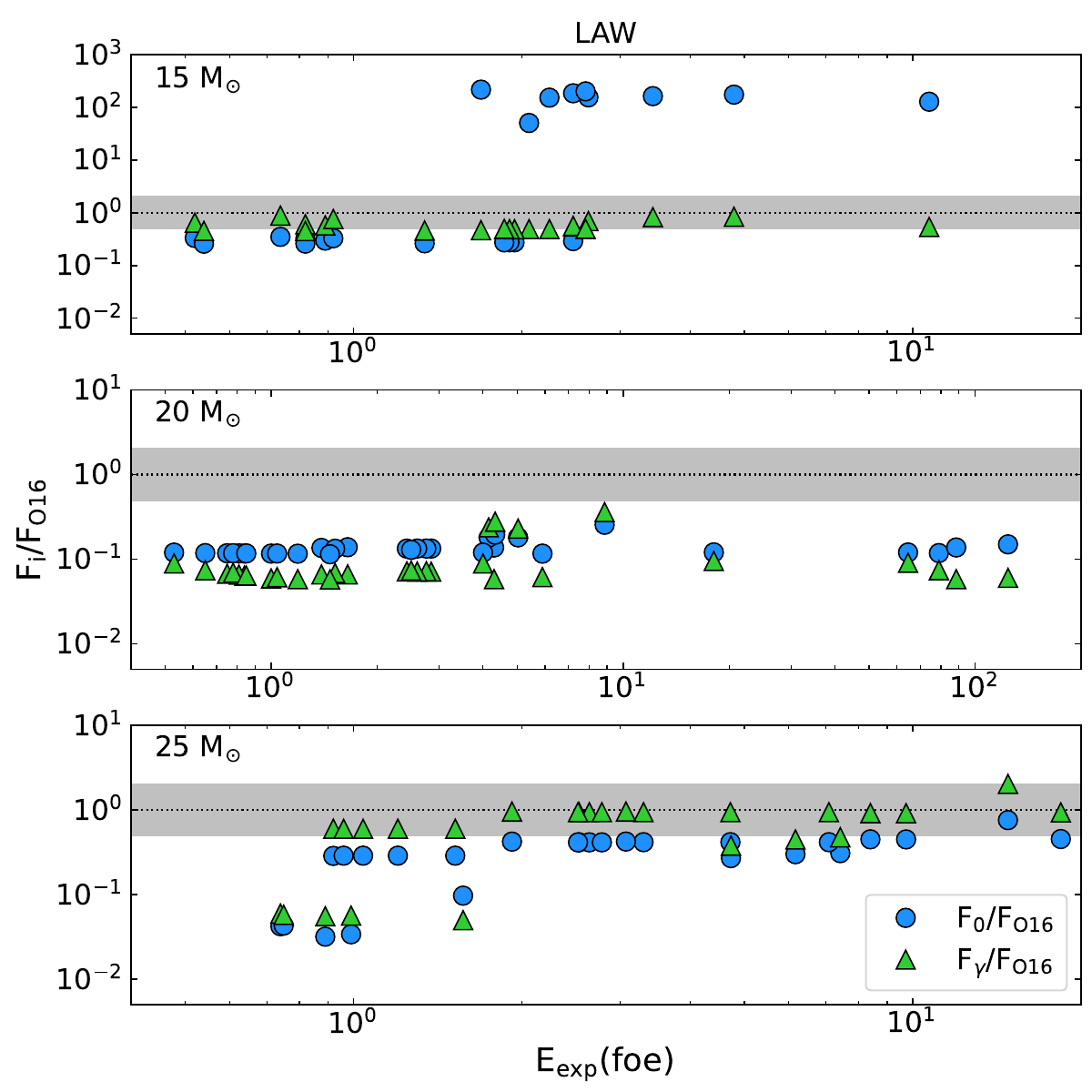}
        \caption{\fz/\fx[O16] (blue diamonds) and \fg/\fx[O16] (green diamonds) in the SBW models as a function of the initial shock velocity (left panels) and in the LAW models as a function of the explosion energy (right panels). The dotted lines represent \fx[i]=\fx[O16], and the shaded gray area represents a factor of 2 above and below the dotted line.}
        \label{fig:f0}
    \end{figure*}   

    Figure \ref{fig:f0} shows the \fz/\fx[O16] and \fg/\fx[O16] ratios in the SBW and LAW sets as a function of the initial mass and of their trend with the explosion energy or, equivalently, the initial shock velocity. In the 20 and 25 \msun\ SBW models, we find that that the increase in the explosion energy produces, on average, an increase within one order of magnitude of the \p-nucleus yields relative to \isotope[16]{O}, in contrast with \cite{choplin:22}. Almost no \p\ nuclei are ejected for low explosion velocities (where the mass cut is larger than the outer mass coordinate for explosive Ne burning). At higher velocities, there is a plateau corresponding to \fz$\sim$\fx[O16], caused by the fact that the external zone of \g-process production coincides with the interface between the CO and He cores, which suppresses the cold \g-process component (see the bottom- and central-left panels of Fig. \ref{fig:spag}, which show the representative cases of \isotope[130]{Ba} and \isotope[196]{Hg}). The production of the hot component is instead boosted by the acceleration of the shock wave in the outer region of the C shell, because the temperature is kept high enough to have a larger \g-process zone (the top-left panel of Fig. \ref{fig:spag} shows the representative case of \isotope[92]{Mo}). The \fg\  increases more than \fz\ because it encompasses fewer isotopes that contribute the most to \fz. The 15 \msun\ SBW models instead experience a C-O shell merger before the explosion. As discussed in Paper I, the yields of all the \p\ nuclei heavier than \isotope[108]{Cd} are almost exclusively dominated by the production obtained in the C-O shell mergers, with a negligible contribution from the explosive nucleosynthesis. Here we confirm this results: we find that, across a wide range of explosion energies, the average overproduction is constant and is equal to or higher than that of \isotope[16]{O} (with \fz/\fx[O16]$\sim1$, \fg/\fx[O16]$\sim2-3$). Moreover, the most produced \p\ nuclei do not change with increasing explosion energy, because the dominant contribution is always produced at the typical temperature of the O-burning shell in the pre-supernova stage. In the whole explosion velocity range, \isotope[130,132]{Ba} and \isotope[144]{Sm} are the nuclei that contribute to \fg. We note, however, that neither the more energetic explosions nor the C-O shell mergers can fully fill the production gap between the Mo-Ru isotopes and the other \p\ nuclei in the SBW models.

    In LAW models the explosion energy, $E_{\rm exp}$, is the hydrodynamic result of a complex interplay between three initial parameters: the amount of injected energy ($E_{\rm inj}$) and the duration ($t_{inj}$) and mass location ($M_{inj}$) of the energy deposition (Sect. \ref{subsec:law}). Only models with $E_{\rm exp} > 2\ \rm foe$ have an efficient production of \p\ nuclei (i.e., \fz/\fx[O16]\ and/or \fg/\fx[O16] $\geq1$). We note, however, that in the 15 \msun\ models the \p\ nucleus production is often dominated by the ejection of some \A-rich freeze-out material rather than the \g-process nucleosynthesis. In nine of the 15 \msun\ models, in fact, we find values of \fz/\fx[O16] as high as 200, but with \fg/\fx[O16] $\leq 1$, due to the extremely large yields of some light \p\ nuclei (from \isotope[74]{Se} to \isotope[94]{Mo}) that are produced very close to the mass cut, just below the abundance peak of \isotope[56]{Ni}, as shown in the top-right panel of Fig. \ref{fig:spag}.
    
    Finally, we find that in LAW models, \fg\ receives the most contribution from intermediate and heavy \p\ nuclei, while in SBW models it receives the most contribution from the light and intermediate \p\ nuclei (see Tables \ref{tab:15}, \ref{tab:20}, and \ref{tab:25}). It follows that the SBW models experience hotter \g-process nucleosynthesis compared to LAW models. This is in qualitative agreement with what was already found by \cite{woosley:95} and \cite{limongi:03a}, who observed that a radiation-dominated technique tends to result in higher peak temperatures than those estimated via a hydrodynamic calculation.

\section{Summary and conclusions} \label{sec:conclusions}

    We studied the effect of a wide range of explosion energies and parameters on the production of \p\ nuclei in CCSNe. We compared the results obtained from a semi-analytical model \citep[SBW;][]{pignatari:16,ritter:18} with those from a set of hydrodynamic simulations \citep[LAW;][]{fryer:18,lawson:22}, using two different definitions of the average \p-nucleus overproduction factors (\fz\ and \fg). As discussed in Paper I \citep[see also][]{farmer:16,BR24}, small differences in the input parameters used by different authors to calculate massive star models with the same initial masses and metallicities (and potentially rotation velocities) can lead to very different structures at the time of the collapse. The propagation of the shock wave and the explosive nucleosynthesis (and therefore the \g\   process) are then influenced by these differences. For this reason, finding that the overall production of \p\ nuclei increases with the increase in the explosion power, regardless of the set and the explosion method used, is a remarkable result.

    In the case of SBW models that do not experience a C-O shell merger (the 20 and 25 \msun\ models), the increase in the explosion energy produces an increase within one order of magnitude of the \p-nucleus yields. Conversely, the \p-nucleus yields from the model that experiences a C-O shell merger during the pre-supernova stage (15 \msun) are not significantly affected by the explosion because the pre-supernova production survives and dominates over the explosive nucleosynthesis for the wide range of explosion velocities we have explored. In any case, neither the increase in the explosion energy nor the occurrence of C-O shell mergers resolve the problem of the \g-process underproduction of Mo and Ru isotopes. 
    
    In the LAW models there is no monotonic trend with explosion energy, but there is an overall increase in the \p-nucleus production in the case of more energetic supernovae. The ejection of some \A-rich freeze-out material, in the case of the explosions above 2 foe in the 15 \msun\ models, increases the Mo-Ru overproduction relative to \isotope[16]{O} by up to two orders of magnitude. Recently, \cite{sasaki:22} suggested that high-energy hypernovae ($\rm E_{exp} \geq 10$ foe) could also provide the right conditions for an efficient $\nu p$ process \citep{froehlich:06}, which may resolve the problem with Mo-Ru isotopes. In any case, high-energy supernovae 
    could represent the key to resolving the long-pending problem about the underproduction of \isotope[92,94]{Mo} and \isotope[96]{Ru}, although not via the \g\  process alone. We further note that the models included in our analysis did not consider 
    neutrino wind nucleosynthesis, and therefore we are not able to estimate the impact at different explosion energies of the $\nu p$ process nor the neutrino spallation contribution, which is crucial for the production of the \p-nucleus $\rm^{138}La$ and $\rm^{180}Ta$ \citep[see][and the detailed description in Paper I]{cumming:85,goriely:01}.
    
    Finally, we stress that with this work we explored only a portion of the wide parameter space that determines the production of \p\ nuclei. For instance, more recent and sophisticated multidimensional simulations of the final stages and explosion of massive stars show some features that are impossible to reproduce in 1D codes. 
    \cite{rizzuti:22} show that a significant entrainment is found in convective Ne-shell burning, confirming that the burning stages that lead to the collapse and to the explosion are still poorly constrained by 1D models. Moreover, most 3D CCSN simulations show that the dynamics of the innermost layers of the ejecta is strongly aspherical, with the formation of structures such as bubbles, plumes, and bullets that may even exceed the speed of the outer material and escape the fallback \citep[][]{wongwathanarat:15,burrows:21}. Asymmetries in the explosion may eventually affect the spatial distribution of the free electron fraction, $Y_{\rm e}$, near the Fe core and, in turn, the production of the lightest \p\ nuclei (up to Pd) via \A-rich freeze-out or the $\nu p$ process \citep[][]{hoffman:96,arcones:13}. A further exploration of the production of \-nuclei in multidimensional simulations will be needed in the future to confirm these predictions. 

\begin{acknowledgements}
    We thank the support from the NKFI via K-project 138031 and the Lend\"ulet Program LP2023-10 of the Hungarian Academy of Sciences. LR and MP acknowledge the support to NuGrid from JINA-CEE (NSF Grant PHY-1430152) and STFC (through the University of Hull’s Consolidated Grant ST/R000840/1), and ongoing access to {\tt viper}, the University of Hull High Performance Computing Facility. LR acknowledges the support from the ChETEC-INFRA -- Transnational Access Projects 22102724-ST and 23103142-ST. This work was supported by the European Union’s Horizon 2020 research and innovation programme (ChETEC-INFRA -- Project no. 101008324), and the IReNA network supported by US NSF AccelNet (Grant No. OISE-1927130). ML was also supported by the NKFIH excellence grant TKP2021-NKTA-64.
\end{acknowledgements}


\begin{appendix}

%
%
%

\section{Tables of average overproduction factors}

    Here we present the different average overproduction factors for all the models analyzed in this work. In particular: 15 \msun\ models in Table \ref{tab:15}; 20 \msun\ in Table \ref{tab:20}; and 25 \msun\ in Table \ref{tab:25}. We repeat below the different definitions for \fz, \fz$\rm ^{I,II,III}$, and \fg\ \citep[see also Table 2 of][]{roberti:23a}. In the case of LAW, we use the same model names as in the online tables (Sect. \ref{subsec:law}). In the case of SBW, the name of each model is composed of the mass in solar masses (M), followed by the absolute metallicity (Z) and the velocity in cm/s (V). The column "$E_{\rm exp}$ {\rm or} $v_{\rm shock}$" contains the final explosion energy in foe (10$^{51}$ erg) in the case of the LAW models, or the initial shock velocity in $\rm 10^9\ cm\ s^{-1}$ in the case of the SBW models.

    \begin{itemize}
        \item \fz: no \p\ nuclei are excluded.
        \item \fz$\rm ^{I}$: the excluded \p\ nuclei are \isotope[74]{Se}, \isotope[78]{Kr}, \isotope[84]{Sr}, \isotope[92,94]{Mo}, \isotope[96,98]{Ru}, \isotope[108]{Cd}, \isotope[113]{In}, \isotope[115]{Sn}, \isotope[138]{La}, \isotope[152]{Gd}, \isotope[164]{Er}, and \isotope[180]{Ta}.   
        \item \fz$\rm ^{II}$: the excluded \p\ nuclei are \isotope[74]{Se}, \isotope[78]{Kr}, \isotope[84]{Sr}, \isotope[92,94]{Mo}, and \isotope[96,98]{Ru}.             
        \item \fz$\rm ^{III}$: the excluded \p\ nuclei are \isotope[108]{Cd}, \isotope[113]{In}, \isotope[115]{Sn}, \isotope[138]{La}, \isotope[152]{Gd}, \isotope[164]{Er}, and \isotope[180]{Ta}.                    
        \item \fg: all \p\ nuclei are excluded except for the three most produced \g-only nuclei, which are listed in the last column. 
    \end{itemize}

   \begin{table*}
      \caption{Average overproduction factors in the 15 \msun\ models.}             
      \label{tab:15}   
      \small
      \centering                          
      \begin{tabular}{lcccccccc}        
      \hline\hline                 
      Model & $E_{\rm exp}$ {\rm or} $v_{\rm shock}$ & \fx[O16] & \fz & $\rm F_0^I$ & $\rm F_0^{II}$ & $\rm F_0^{III}$ & $\rm F_{\gamma}$ & $\rm F_{\gamma}$ isotopes \\  
      \hline
      \multicolumn{9}{c}{LAW}\\
      \hline  
      M15s$\_$f1$\_$216M1.3aal    &   2.63    &  1.21e+01  &  1.87e+03  &  4.70e+00  &  4.47e+00  &  2.34e+03  &  8.37e+00  &  \isotope[190]{Pt}, \isotope[130]{Ba}, \isotope[158]{Dy} \\ 
      M15s$\_$f1$\_$216M1.3abl    &   4.79    &  1.19e+01  &  2.07e+03  &  5.41e+00  &  5.01e+00  &  2.59e+03  &  9.98e+00  &  \isotope[158]{Dy}, \isotope[190]{Pt}, \isotope[196]{Hg} \\ 
      M15s$\_$f1$\_$216M1.3bdl    &   1.69    &  1.22e+01  &  2.64e+03  &  3.44e+00  &  3.61e+00  &  3.30e+03  &  5.68e+00  &  \isotope[184]{Os}, \isotope[130]{Ba}, \isotope[158]{Dy} \\ 
      M15s$\_$f1$\_$216M1.3bel    &  10.70    &  1.23e+01  &  1.57e+03  &  4.01e+00  &  3.94e+00  &  1.96e+03  &  6.59e+00  &  \isotope[130]{Ba}, \isotope[158]{Dy}, \isotope[190]{Pt} \\ 
      M15s$\_$f1$\_$216M1.3bfl    &   0.89    &  1.20e+01  &  3.55e+00  &  3.91e+00  &  3.85e+00  &  3.52e+00  &  6.90e+00  &  \isotope[158]{Dy}, \isotope[190]{Pt}, \isotope[130]{Ba} \\ 
      M15s$\_$f1$\_$216M1.3bgl    &   0.92    &  1.26e+01  &  4.13e+00  &  4.87e+00  &  4.72e+00  &  4.10e+00  &  9.56e+00  &  \isotope[190]{Pt}, \isotope[196]{Hg}, \isotope[158]{Dy} \\ 
      M15s$\_$f1$\_$216M1.3bhl    &   0.74    &  1.26e+01  &  4.39e+00  &  5.31e+00  &  5.04e+00  &  4.44e+00  &  1.11e+01  &  \isotope[190]{Pt}, \isotope[196]{Hg}, \isotope[158]{Dy} \\ 
      M15s$\_$f1$\_$216M1.3bil    &   0.82    &  1.12e+01  &  3.59e+00  &  3.98e+00  &  3.84e+00  &  3.64e+00  &  6.76e+00  &  \isotope[158]{Dy}, \isotope[130]{Ba}, \isotope[190]{Pt} \\ 
      M15s$\_$f1$\_$216M1.3bjl    &   0.52    &  1.08e+01  &  3.59e+00  &  4.05e+00  &  3.83e+00  &  3.69e+00  &  7.00e+00  &  \isotope[130]{Ba}, \isotope[190]{Pt}, \isotope[158]{Dy} \\ 
      M15s$\_$f1$\_$216M1.3bkl    &   0.30    &  1.05e+01  &  3.66e+00  &  4.12e+00  &  3.87e+00  &  3.79e+00  &  7.68e+00  &  \isotope[158]{Dy}, \isotope[190]{Pt}, \isotope[130]{Ba} \\ 
      M15s$\_$f1$\_$216M1.3sl     &   1.34    &  1.26e+01  &  3.31e+00  &  3.51e+00  &  3.69e+00  &  3.09e+00  &  5.79e+00  &  \isotope[158]{Dy}, \isotope[130]{Ba}, \isotope[190]{Pt} \\ 
      M15s$\_$f1$\_$216M1.3tl     &   0.82    &  1.28e+01  &  3.32e+00  &  3.48e+00  &  3.69e+00  &  3.07e+00  &  5.73e+00  &  \isotope[130]{Ba}, \isotope[158]{Dy}, \isotope[190]{Pt} \\ 
      M15s$\_$f1$\_$216M1.3vl     &   0.34    &  1.25e+01  &  3.27e+00  &  3.41e+00  &  3.68e+00  &  2.96e+00  &  5.75e+00  &  \isotope[158]{Dy}, \isotope[130]{Ba}, \isotope[190]{Pt} \\ 
      M15s$\_$f1$\_$216M1.3xl     &   0.54    &  1.28e+01  &  3.31e+00  &  3.45e+00  &  3.69e+00  &  3.04e+00  &  5.79e+00  &  \isotope[158]{Dy}, \isotope[130]{Ba}, \isotope[190]{Pt} \\ 
      M15s$\_$f1$\_$216M1.3yl     &   2.47    &  1.22e+01  &  3.53e+00  &  3.87e+00  &  3.90e+00  &  3.42e+00  &  6.70e+00  &  \isotope[158]{Dy}, \isotope[130]{Ba}, \isotope[190]{Pt} \\ 
      M15s$\_$f1$\_$216M1.3zl     &   2.47    &  1.16e+01  &  2.14e+03  &  3.78e+00  &  3.75e+00  &  2.68e+03  &  6.52e+00  &  \isotope[158]{Dy}, \isotope[130]{Ba}, \isotope[190]{Pt} \\ 
      M15s$\_$t1                  &   2.06    &  1.22e+01  &  6.16e+02  &  3.49e+00  &  3.63e+00  &  7.69e+02  &  5.96e+00  &  \isotope[158]{Dy}, \isotope[130]{Ba}, \isotope[190]{Pt} \\ 
      M15s$\_$t2                  &   1.94    &  1.23e+01  &  3.38e+00  &  3.50e+00  &  3.64e+00  &  3.21e+00  &  5.98e+00  &  \isotope[158]{Dy}, \isotope[130]{Ba}, \isotope[190]{Pt} \\ 
      M15s$\_$t3                  &   1.90    &  1.23e+01  &  3.38e+00  &  3.51e+00  &  3.65e+00  &  3.21e+00  &  6.01e+00  &  \isotope[158]{Dy}, \isotope[130]{Ba}, \isotope[190]{Pt} \\ 
      M15s$\_$t5                  &   1.86    &  1.23e+01  &  3.37e+00  &  3.52e+00  &  3.66e+00  &  3.20e+00  &  6.06e+00  &  \isotope[158]{Dy}, \isotope[130]{Ba}, \isotope[190]{Pt} \\ 
      M15s$\_$t6                  &   2.24    &  1.22e+01  &  1.86e+03  &  3.50e+00  &  3.64e+00  &  2.32e+03  &  5.98e+00  &  \isotope[158]{Dy}, \isotope[130]{Ba}, \isotope[190]{Pt} \\ 
      M15s$\_$t8                  &   2.60    &  1.21e+01  &  2.43e+03  &  3.42e+00  &  3.56e+00  &  3.04e+03  &  5.96e+00  &  \isotope[158]{Dy}, \isotope[130]{Ba}, \isotope[184]{Os} \\ 
      M15s$\_$t9                  &   3.43    &  1.21e+01  &  1.97e+03  &  4.99e+00  &  4.75e+00  &  2.47e+03  &  9.99e+00  &  \isotope[190]{Pt}, \isotope[158]{Dy}, \isotope[196]{Hg} \\ 
      \hline
      \multicolumn{9}{c}{SBW}\\
      \hline 
      M15.0Z2.0e-02.V2.0e08$^a$   &  0.40     &  1.49E+01  &  1.51E+01  &  1.82E+01  &  1.49E+01  &  1.76E+01  &  4.44E+01  &  \isotope[132]{Ba}, \isotope[130]{Ba}, \isotope[144]{Sm} \\  
      M15.0Z2.0e-02.V4.0e08$^a$   &  0.80     &  1.49E+01  &  1.51E+01  &  1.82E+01  &  1.49E+01  &  1.76E+01  &  4.44E+01  &  \isotope[132]{Ba}, \isotope[130]{Ba}, \isotope[144]{Sm} \\
      M15.0Z2.0e-02.V1.0e09$^a$   &  1.00     &  1.47E+01  &  1.51E+01  &  1.75E+01  &  1.42E+01  &  1.77E+01  &  4.20E+01  &  \isotope[132]{Ba}, \isotope[130]{Ba}, \isotope[144]{Sm} \\
      M15.0Z2.0e-02.V1.3e09$^a$   &  1.33     &  1.44E+01  &  1.48E+01  &  1.69E+01  &  1.38E+01  &  1.74E+01  &  4.02E+01  &  \isotope[132]{Ba}, \isotope[130]{Ba}, \isotope[144]{Sm} \\
      M15.0Z2.0e-02.V2.0e09$^a$   &  2.00     &  1.38E+01  &  1.43E+01  &  1.60E+01  &  1.31E+01  &  1.68E+01  &  3.71E+01  &  \isotope[132]{Ba}, \isotope[130]{Ba}, \isotope[144]{Sm} \\
      M15.0Z2.0e-02.V3.0e09$^a$   &  3.00     &  1.31E+01  &  1.38E+01  &  1.46E+01  &  1.21E+01  &  1.61E+01  &  3.26E+01  &  \isotope[132]{Ba}, \isotope[130]{Ba}, \isotope[144]{Sm} \\
      M15.0Z2.0e-02.V4.0e09$^a$   &  4.00     &  1.24E+01  &  1.34E+01  &  1.37E+01  &  1.16E+01  &  1.54E+01  &  2.96E+01  &  \isotope[132]{Ba}, \isotope[130]{Ba}, \isotope[144]{Sm} \\
      M15.0Z2.0e-02.V5.0e09$^a$   &  5.00     &  1.23E+01  &  1.32E+01  &  1.36E+01  &  1.15E+01  &  1.52E+01  &  2.94E+01  &  \isotope[132]{Ba}, \isotope[130]{Ba}, \isotope[144]{Sm} \\
      \hline                                   
      \end{tabular}\\
      \tablefoot{$^a$These models experience a C--O shell merger.}
   \end{table*}

    \begin{table*}
      \caption{Average overproduction factors in the 20 \msun\ models.}             
      \label{tab:20}   
      \small
      \centering                          
      \begin{tabular}{lcccccccc}        
      \hline\hline   
      Model & $E_{\rm exp}$ {\rm or} $v_{\rm shock}$ & \fx[O16] & \fz & $\rm F_0^I$ & $\rm F_0^{II}$ & $\rm F_0^{III}$ & $\rm F_{\gamma}$ & $\rm F_{\gamma}$ isotopes \\  
      \hline
      \multicolumn{9}{c}{LAW}\\
      \hline                         
            M20s$\_$f1$\_$240M1.47al  &    2.85   &  2.53e+01  &  3.38e+00  &  1.26e+00  &  3.61e+00  &  1.56e+00  &  1.80e+00  &  \isotope[158]{Dy}, \isotope[114]{Sn}, \isotope[180]{ W} \\  
            M20s$\_$f1$\_$240M1.47bl  &    5.03   &  2.50e+01  &  4.52e+00  &  3.13e+00  &  5.09e+00  &  2.91e+00  &  5.73e+00  &  \isotope[158]{Dy}, \isotope[184]{Os}, \isotope[196]{Hg} \\  
            M20s$\_$f1$\_$240M1.47cl  &    8.86   &  2.47e+01  &  6.30e+00  &  4.89e+00  &  6.29e+00  &  5.26e+00  &  8.89e+00  &  \isotope[130]{Ba}, \isotope[132]{Ba}, \isotope[158]{Dy} \\  
            M20s$\_$f1$\_$240M1.47dl  &    1.65   &  2.57e+01  &  3.56e+00  &  1.17e+00  &  3.51e+00  &  1.83e+00  &  1.70e+00  &  \isotope[158]{Dy}, \isotope[114]{Sn}, \isotope[180]{ W} \\  
            M20s$\_$f1$\_$240M1.47el  &    1.00   &  2.36e+01  &  2.75e+00  &  8.92e-01  &  3.20e+00  &  9.03e-01  &  1.40e+00  &  \isotope[114]{Sn}, \isotope[158]{Dy}, \isotope[180]{ W} \\  
            M20s$\_$f1$\_$240M1.47fl  &    0.84   &  2.16e+01  &  2.52e+00  &  9.07e-01  &  2.92e+00  &  9.19e-01  &  1.38e+00  &  \isotope[114]{Sn}, \isotope[158]{Dy}, \isotope[180]{ W} \\  
            M20s$\_$f1$\_$240M1.47gl  &    0.75   &  2.05e+01  &  2.41e+00  &  9.14e-01  &  2.77e+00  &  9.28e-01  &  1.38e+00  &  \isotope[114]{Sn}, \isotope[158]{Dy}, \isotope[180]{ W} \\  
            M20s$\_$f1$\_$240M1.47hl  &    2.76   &  2.53e+01  &  3.38e+00  &  1.27e+00  &  3.64e+00  &  1.54e+00  &  1.82e+00  &  \isotope[158]{Dy}, \isotope[114]{Sn}, \isotope[180]{ W} \\  
         M20s$\_$f1$\_$240M1.47redal  &  124.00   &  2.61e+01  &  3.91e+00  &  1.09e+00  &  3.45e+00  &  2.26e+00  &  1.57e+00  &  \isotope[114]{Sn}, \isotope[102]{Pd}, \isotope[180]{ W} \\  
         M20s$\_$f1$\_$240M1.47redbl  &   64.50   &  1.49e+01  &  1.79e+00  &  9.55e-01  &  1.98e+00  &  9.75e-01  &  1.35e+00  &  \isotope[158]{Dy}, \isotope[114]{Sn}, \isotope[102]{Pd} \\  
         M20s$\_$f1$\_$240M1.47redcl  &    5.90   &  2.26e+01  &  2.63e+00  &  8.98e-01  &  3.05e+00  &  9.10e-01  &  1.39e+00  &  \isotope[114]{Sn}, \isotope[158]{Dy}, \isotope[180]{ W} \\  
         M20s$\_$f1$\_$240M1.47reddl  &   18.10   &  1.42e+01  &  1.71e+00  &  9.61e-01  &  1.88e+00  &  9.81e-01  &  1.35e+00  &  \isotope[158]{Dy}, \isotope[114]{Sn}, \isotope[102]{Pd} \\  
         M20s$\_$f1$\_$240M1.47redel  &   78.90   &  1.82e+01  &  2.15e+00  &  9.29e-01  &  2.44e+00  &  9.46e-01  &  1.35e+00  &  \isotope[114]{Sn}, \isotope[158]{Dy}, \isotope[180]{ W} \\  
         M20s$\_$f1$\_$240M1.47redfl  &   88.40   &  2.56e+01  &  3.52e+00  &  9.78e-01  &  3.38e+00  &  1.75e+00  &  1.49e+00  &  \isotope[114]{Sn}, \isotope[102]{Pd}, \isotope[158]{Dy} \\  
         M20s$\_$f1$\_$240M1.47redgl  &    4.30   &  2.56e+01  &  3.52e+00  &  9.79e-01  &  3.38e+00  &  1.75e+00  &  1.49e+00  &  \isotope[114]{Sn}, \isotope[102]{Pd}, \isotope[158]{Dy} \\  
            M20s$\_$f1$\_$300M1.56al  &    4.15   &  2.52e+01  &  4.53e+00  &  3.14e+00  &  5.12e+00  &  2.91e+00  &  6.00e+00  &  \isotope[158]{Dy}, \isotope[184]{Os}, \isotope[196]{Hg} \\  
            M20s$\_$f1$\_$300M1.56bl  &    2.43   &  2.55e+01  &  3.40e+00  &  1.27e+00  &  3.64e+00  &  1.56e+00  &  1.83e+00  &  \isotope[158]{Dy}, \isotope[114]{Sn}, \isotope[180]{ W} \\  
            M20s$\_$f1$\_$300M1.56cl  &    1.39   &  2.56e+01  &  3.48e+00  &  1.17e+00  &  3.53e+00  &  1.69e+00  &  1.69e+00  &  \isotope[158]{Dy}, \isotope[114]{Sn}, \isotope[180]{ W} \\  
            M20s$\_$f1$\_$300M1.56dl  &    0.81   &  2.10e+01  &  2.46e+00  &  9.11e-01  &  2.83e+00  &  9.24e-01  &  1.38e+00  &  \isotope[114]{Sn}, \isotope[158]{Dy}, \isotope[180]{ W} \\  
            M20s$\_$f1$\_$300M1.56el  &    0.65   &  1.84e+01  &  2.17e+00  &  9.30e-01  &  2.46e+00  &  9.46e-01  &  1.36e+00  &  \isotope[114]{Sn}, \isotope[158]{Dy}, \isotope[180]{ W} \\  
            M20s$\_$f1$\_$300M1.56gl  &    4.33   &  2.53e+01  &  4.93e+00  &  3.70e+00  &  5.55e+00  &  3.38e+00  &  7.01e+00  &  \isotope[130]{Ba}, \isotope[158]{Dy}, \isotope[132]{Ba} \\  
            M20s$\_$f1$\_$300M1.56hl  &    2.60   &  2.56e+01  &  3.42e+00  &  1.29e+00  &  3.69e+00  &  1.55e+00  &  1.83e+00  &  \isotope[158]{Dy}, \isotope[114]{Sn}, \isotope[180]{ W} \\  
            M20s$\_$f1$\_$300M1.56il  &    1.52   &  2.55e+01  &  3.39e+00  &  1.18e+00  &  3.55e+00  &  1.57e+00  &  1.75e+00  &  \isotope[158]{Dy}, \isotope[114]{Sn}, \isotope[180]{ W} \\  
            M20s$\_$f1$\_$300M1.56jl  &    1.19   &  2.41e+01  &  2.80e+00  &  8.88e-01  &  3.27e+00  &  8.98e-01  &  1.40e+00  &  \isotope[114]{Sn}, \isotope[158]{Dy}, \isotope[180]{ W} \\  
            M20s$\_$f1$\_$300M1.56kl  &    1.04   &  2.27e+01  &  2.65e+00  &  8.98e-01  &  3.08e+00  &  9.10e-01  &  1.39e+00  &  \isotope[114]{Sn}, \isotope[158]{Dy}, \isotope[180]{ W} \\  
            M20s$\_$f1$\_$300M1.56ll  &    0.78   &  1.98e+01  &  2.33e+00  &  9.20e-01  &  2.67e+00  &  9.34e-01  &  1.37e+00  &  \isotope[114]{Sn}, \isotope[158]{Dy}, \isotope[180]{ W} \\  
          M20s$\_$f1$\_$300M1.56t2al  &    2.50   &  2.56e+01  &  3.31e+00  &  1.32e+00  &  3.53e+00  &  1.59e+00  &  1.88e+00  &  \isotope[158]{Dy}, \isotope[114]{Sn}, \isotope[180]{ W} \\  
          M20s$\_$f1$\_$300M1.56t2bl  &    1.47   &  2.44e+01  &  2.79e+00  &  8.88e-01  &  3.25e+00  &  8.97e-01  &  1.41e+00  &  \isotope[114]{Sn}, \isotope[158]{Dy}, \isotope[180]{ W} \\  
          M20s$\_$f1$\_$300M1.56t2cl  &    0.85   &  2.16e+01  &  2.53e+00  &  9.07e-01  &  2.92e+00  &  9.19e-01  &  1.38e+00  &  \isotope[114]{Sn}, \isotope[158]{Dy}, \isotope[180]{ W} \\  
          M20s$\_$f1$\_$300M1.56t2dl  &    0.53   &  1.52e+01  &  1.82e+00  &  9.53e-01  &  2.02e+00  &  9.72e-01  &  1.35e+00  &  \isotope[158]{Dy}, \isotope[114]{Sn}, \isotope[102]{Pd} \\  
          M20s$\_$f1$\_$300M1.56t2el  &    4.00   &  1.52e+01  &  1.82e+00  &  9.53e-01  &  2.02e+00  &  9.72e-01  &  1.35e+00  &  \isotope[158]{Dy}, \isotope[114]{Sn}, \isotope[102]{Pd} \\ 
      \hline
      \multicolumn{9}{c}{SBW}\\
      \hline 
      M20.0Z2.0e-02.V2.0e08           &  0.40     &  1.81E+01 &  1.31E+00   &  9.17E-01  &  1.40E+00  &  9.25E-01  &  1.47E+00    & \isotope[114]{Sn}, \isotope[158]{Dy}, \isotope[102]{Pd} \\    
      M20.0Z2.0e-02.V4.0e08           &  0.80     &  1.78E+01 &  1.23E+00   &  8.92E-01  &  1.31E+00  &  8.97E-01  &  1.38E+00    & \isotope[114]{Sn}, \isotope[158]{Dy}, \isotope[168]{Yb} \\    
      M20.0Z2.0e-02.V1.0e09           &  1.00     &  1.80E+01 &  2.07E+00   &  1.68E+00  &  2.37E+00  &  1.48E+00  &  4.44E+00    & \isotope[184]{Os}, \isotope[180]{ W}, \isotope[190]{Pt} \\    
      M20.0Z2.0e-02.V1.3e09           &  1.33     &  1.86E+01 &  7.07E+00   &  9.27E+00  &  8.35E+00  &  7.44E+00  &  2.05E+01    & \isotope[130]{Ba}, \isotope[158]{Dy}, \isotope[196]{Hg} \\    
      M20.0Z2.0e-02.V2.0e09           &  2.00     &  1.94E+01 &  1.40E+01   &  1.31E+01  &  1.15E+01  &  1.59E+01  &  2.48E+01    & \isotope[130]{Ba}, \isotope[158]{Dy}, \isotope[196]{Hg} \\    
      M20.0Z2.0e-02.V3.0e09           &  3.00     &  1.77E+01 &  1.99E+01   &  2.22E+01  &  1.78E+01  &  2.37E+01  &  4.76E+01    & \isotope[184]{Os}, \isotope[190]{Pt}, \isotope[196]{Hg} \\    
      M20.0Z2.0e-02.V4.0e09           &  4.00     &  1.46E+01 &  1.75E+01   &  1.58E+01  &  1.30E+01  &  2.07E+01  &  5.04E+01    & \isotope[130]{Ba}, \isotope[144]{Sm}, \isotope[124]{Xe} \\    
      M20.0Z2.0e-02.V5.0e09           &  5.00     &  1.29E+01 &  1.46E+01   &  8.47E+00  &  7.12E+00  &  1.75E+01  &  2.18E+01    & \isotope[102]{Pd}, \isotope[130]{Ba}, \isotope[106]{Cd} \\ 
      \hline                                   
      \end{tabular}\\
   \end{table*}

    \begin{table*}
      \caption{Average overproduction factors in the 25 \msun\ models.}              
      \label{tab:25}   
      \small
      \centering                          
      \begin{tabular}{lcccccccc}        
      \hline\hline   
      Model & $E_{\rm exp}$ {\rm or} $v_{\rm shock}$ & \fx[O16] & \fz & $\rm F_0^I$ & $\rm F_0^{II}$ & $\rm F_0^{III}$ & $\rm F_{\gamma}$ & $\rm F_{\gamma}$ isotopes \\  
      \hline
      \multicolumn{9}{c}{LAW}\\
      \hline                         
       M25s$\_$f1$\_$280M1.83al    &     4.73  &  4.98e+01  &  1.34e+01  &  9.46e+00  &  1.38e+01  &  1.01e+01  &  1.88e+01  &  \isotope[130]{Ba}, \isotope[184]{Os}, \isotope[158]{Dy}\\
       M25s$\_$f1$\_$280M1.83bl    &     6.17  &  5.01e+01  &  1.50e+01  &  1.11e+01  &  1.53e+01  &  1.18e+01  &  2.24e+01  &  \isotope[184]{Os}, \isotope[130]{Ba}, \isotope[158]{Dy}\\
       M25s$\_$f1$\_$280M1.83cl    &    14.80  &  5.13e+01  &  3.88e+01  &  4.85e+01  &  4.28e+01  &  4.21e+01  &  1.04e+02  &  \isotope[130]{Ba}, \isotope[158]{Dy}, \isotope[132]{Ba}\\
       M25s$\_$f1$\_$280M1.83dl    &     7.42  &  5.00e+01  &  1.54e+01  &  1.16e+01  &  1.57e+01  &  1.22e+01  &  2.37e+01  &  \isotope[184]{Os}, \isotope[130]{Ba}, \isotope[158]{Dy}\\
       M25s$\_$f1$\_$280M1.83el    &     1.57  &  4.62e+01  &  4.48e+00  &  7.99e-01  &  5.43e+00  &  7.70e-01  &  2.30e+00  &  \isotope[180]{ W}, \isotope[114]{Sn}, \isotope[158]{Dy}\\
       M25s$\_$f1$\_$280M1.83fl    &     0.99  &  3.57e+01  &  1.21e+00  &  8.37e-01  &  1.31e+00  &  8.27e-01  &  2.01e+00  &  \isotope[180]{ W}, \isotope[114]{Sn}, \isotope[158]{Dy}\\
       M25s$\_$f1$\_$280M1.83hl    &     0.74  &  2.89e+01  &  1.22e+00  &  8.47e-01  &  1.31e+00  &  8.53e-01  &  1.72e+00  &  \isotope[180]{ W}, \isotope[158]{Dy}, \isotope[114]{Sn}\\
       M25s$\_$f1$\_$280M1.83ll    &     8.40  &  5.46e+01  &  2.45e+01  &  2.62e+01  &  2.56e+01  &  2.46e+01  &  4.99e+01  &  \isotope[130]{Ba}, \isotope[184]{Os}, \isotope[158]{Dy}\\
       M25s$\_$f1$\_$280M1.83ml    &     9.73  &  5.45e+01  &  2.44e+01  &  2.61e+01  &  2.56e+01  &  2.45e+01  &  4.97e+01  &  \isotope[130]{Ba}, \isotope[184]{Os}, \isotope[158]{Dy}\\
       M25s$\_$f1$\_$280M1.83nl    &    18.40  &  5.44e+01  &  2.46e+01  &  2.65e+01  &  2.59e+01  &  2.48e+01  &  5.08e+01  &  \isotope[130]{Ba}, \isotope[184]{Os}, \isotope[158]{Dy}\\
      M25s$\_$f1$\_$280M1.83ral    &     4.72  &  5.42e+01  &  2.25e+01  &  2.64e+01  &  2.54e+01  &  2.25e+01  &  5.09e+01  &  \isotope[130]{Ba}, \isotope[184]{Os}, \isotope[158]{Dy}\\
      M25s$\_$f1$\_$280M1.83rbl    &     2.53  &  5.42e+01  &  2.24e+01  &  2.64e+01  &  2.54e+01  &  2.25e+01  &  5.09e+01  &  \isotope[130]{Ba}, \isotope[184]{Os}, \isotope[158]{Dy}\\
      M25s$\_$f1$\_$280M1.83rcl    &     3.30  &  5.42e+01  &  2.25e+01  &  2.64e+01  &  2.54e+01  &  2.25e+01  &  5.09e+01  &  \isotope[130]{Ba}, \isotope[184]{Os}, \isotope[158]{Dy}\\
      M25s$\_$f1$\_$280M1.83rdl    &     7.08  &  5.42e+01  &  2.25e+01  &  2.64e+01  &  2.54e+01  &  2.25e+01  &  5.09e+01  &  \isotope[130]{Ba}, \isotope[184]{Os}, \isotope[158]{Dy}\\
      M25s$\_$f1$\_$280M1.83rel    &     2.64  &  5.42e+01  &  2.24e+01  &  2.64e+01  &  2.53e+01  &  2.25e+01  &  5.09e+01  &  \isotope[130]{Ba}, \isotope[184]{Os}, \isotope[158]{Dy}\\
      M25s$\_$f1$\_$280M1.83rfl    &     2.52  &  5.42e+01  &  2.24e+01  &  2.64e+01  &  2.53e+01  &  2.25e+01  &  5.09e+01  &  \isotope[130]{Ba}, \isotope[184]{Os}, \isotope[158]{Dy}\\
      M25s$\_$f1$\_$280M1.83rgl    &     2.78  &  5.42e+01  &  2.24e+01  &  2.64e+01  &  2.53e+01  &  2.25e+01  &  5.09e+01  &  \isotope[130]{Ba}, \isotope[184]{Os}, \isotope[158]{Dy}\\
      M25s$\_$f1$\_$280M1.83rhl    &     0.75  &  2.70e+01  &  1.16e+00  &  8.22e-01  &  1.23e+00  &  8.32e-01  &  1.55e+00  &  \isotope[180]{ W}, \isotope[158]{Dy}, \isotope[114]{Sn}\\
      M25s$\_$f1$\_$280M1.83ril    &     3.07  &  5.14e+01  &  2.17e+01  &  2.55e+01  &  2.44e+01  &  2.18e+01  &  4.91e+01  &  \isotope[130]{Ba}, \isotope[184]{Os}, \isotope[158]{Dy}\\
      M25s$\_$f1$\_$280M1.83rjl    &     1.92  &  5.15e+01  &  2.17e+01  &  2.55e+01  &  2.44e+01  &  2.18e+01  &  4.92e+01  &  \isotope[130]{Ba}, \isotope[184]{Os}, \isotope[158]{Dy}\\
      M25s$\_$f1$\_$280M1.83rnl    &     0.89  &  3.81e+01  &  1.21e+00  &  8.35e-01  &  1.32e+00  &  8.19e-01  &  2.11e+00  &  \isotope[180]{ W}, \isotope[114]{Sn}, \isotope[158]{Dy}\\
      M25s$\_$f1$\_$280M1.83rol    &     0.92  &  5.18e+01  &  1.48e+01  &  1.50e+01  &  1.62e+01  &  1.36e+01  &  3.10e+01  &  \isotope[130]{Ba}, \isotope[184]{Os}, \isotope[158]{Dy}\\
      M25s$\_$f1$\_$280M1.83rpl    &     0.96  &  5.18e+01  &  1.49e+01  &  1.50e+01  &  1.62e+01  &  1.36e+01  &  3.10e+01  &  \isotope[130]{Ba}, \isotope[184]{Os}, \isotope[158]{Dy}\\
      M25s$\_$f1$\_$280M1.83rql    &     1.04  &  5.18e+01  &  1.49e+01  &  1.50e+01  &  1.63e+01  &  1.36e+01  &  3.10e+01  &  \isotope[130]{Ba}, \isotope[184]{Os}, \isotope[158]{Dy}\\
      M25s$\_$f1$\_$280M1.83rrl    &     1.20  &  5.18e+01  &  1.49e+01  &  1.50e+01  &  1.62e+01  &  1.36e+01  &  3.10e+01  &  \isotope[130]{Ba}, \isotope[184]{Os}, \isotope[158]{Dy}\\
      M25s$\_$f1$\_$280M1.83rsl    &     1.52  &  5.18e+01  &  1.49e+01  &  1.50e+01  &  1.62e+01  &  1.36e+01  &  3.10e+01  &  \isotope[130]{Ba}, \isotope[184]{Os}, \isotope[158]{Dy}\\
      \hline
      \multicolumn{9}{c}{SBW}\\
      \hline 
      M25Z2.0e-02.V2.0e08          &  0.40     &  2.00E+01  &  2.07E+00  &  7.52E-01  &  2.36E+00  &  7.87E-01  &  1.10E+00  &  \isotope[114]{Sn}, \isotope[158]{Dy}, \isotope[112]{Sn}\\  
      M25Z2.0e-02.V4.0e08          &  0.80     &  1.96E+01  &  1.96E+00  &  7.19E-01  &  2.24E+00  &  7.45E-01  &  9.99E-01  &  \isotope[114]{Sn}, \isotope[158]{Dy}, \isotope[102]{Pd}\\  
      M25Z2.0e-02.V1.0e09          &  1.00     &  1.93E+01  &  1.83E+00  &  6.77E-01  &  2.10E+00  &  6.92E-01  &  8.72E-01  &  \isotope[114]{Sn}, \isotope[158]{Dy}, \isotope[168]{Yb}\\  
      M25Z2.0e-02.V1.3e09          &  1.33     &  1.93E+01  &  1.83E+00  &  6.78E-01  &  2.11E+00  &  6.92E-01  &  8.72E-01  &  \isotope[114]{Sn}, \isotope[158]{Dy}, \isotope[168]{Yb}\\  
      M25Z2.0e-02.V2.0e09          &  2.00     &  1.94E+01  &  1.75E+00  &  6.78E-01  &  2.00E+00  &  6.92E-01  &  8.72E-01  &  \isotope[114]{Sn}, \isotope[158]{Dy}, \isotope[168]{Yb}\\  
      M25Z2.0e-02.V3.0e09          &  3.00     &  2.00E+01  &  3.79E+00  &  2.26E+00  &  4.56E+00  &  1.88E+00  &  1.11E+01  &  \isotope[180]{ W}, \isotope[184]{Os}, \isotope[174]{Hf}\\  
      M25Z2.0e-02.V4.0e09          &  4.00     &  2.12E+01  &  2.07E+01  &  3.15E+01  &  2.55E+01  &  2.40E+01  &  9.73E+01  &  \isotope[158]{Dy}, \isotope[132]{Ba}, \isotope[196]{Hg}\\  
      M25Z2.0e-02.V5.0e09          &  5.00     &  2.13E+01  &  1.94E+01  &  2.85E+01  &  2.26E+01  &  2.29E+01  &  7.16E+01  &  \isotope[130]{Ba}, \isotope[144]{Sm}, \isotope[132]{Ba}\\  
      \hline                                   
      \end{tabular}\\
   \end{table*}

\end{appendix}

\end{document}